\documentclass[]{revtex4}

\usepackage{ccaption}
\usepackage{times}
\usepackage{amstext}
\usepackage{url}
\usepackage{amssymb}

\usepackage[pdftex]{graphicx}

\begin{document} 

\title{Detecting inertial effects with airborne matter-wave interferometry}

\author{R. Geiger$^{1,4}$, V. M\'enoret$^{1}$, G. Stern$^{1,2,4}$,  N. Zahzam$^{3}$, P. Cheinet$^{1}$,  B. Battelier$^{1,2}$, A. Villing$^{1}$,
F. Moron$^{1}$, M. Lours$^{2}$, Y. Bidel$^{3}$,  A. Bresson$^{3}$, A. Landragin$^{2}$ \& P. Bouyer$^{1,5}$}

\affiliation{$^{1}$Laboratoire Charles Fabry, UMR 8501, Institut d'Optique, CNRS, Univ. Paris Sud 11, 2, Avenue Augustin Fresnel, 91127 Palaiseau, France} 
\affiliation{$^{2}$LNE-SYRTE, Observatoire de Paris, CNRS and UPMC, 61 avenue de l'Observatoire, 75014 Paris, France}
\affiliation{$^{3}$ONERA, DMPH, Chemin de la Huni\`ere, 91761 Palaiseau, France}
\affiliation{$^{4}$CNES, 18 Avenue Edouard Belin, 31401 Toulouse, France}
\affiliation{$^{5}$Laboratoire Photonique Num\'erique et Nanosciences, Universit\'e Bordeaux 1, IOGS and CNRS, 351 cours de la Lib\'eration, 33405 Talence, France}

\begin{abstract}
Inertial sensors relying on atom interferometry offer a breakthrough advance in a variety of applications, such as inertial navigation,  gravimetry or ground and space based tests of fundamental physics.
These instruments require a quiet environment to reach their performance and  using them outside the laboratory  remains a challenge.
Here, we report the first operation of an airborne matter wave accelerometer set up aboard a 0-g plane and operating during the standard gravity (1-g)  and microgravity (0-g) phases of the flight.
At 1-g, the sensor can detect inertial effects more than 300 times weaker than the typical acceleration fluctuations of the aircraft.
We describe the improvement of the interferometer sensitivity in 0-g, which reaches  $2\times 10^{-4} \ \text{m.s}^{-2}/\sqrt{\text{Hz}}$ with our current setup.
We finally discuss the extension of our method to airborne and spaceborne tests of the Universality of Free Fall with matter waves.
\end{abstract}

\maketitle

Matter wave inertial sensing relies on the capability of controling the wave nature of matter to build an interferometer and accurately measure a phase difference \cite{Berman,Cronin}.
Since the particle associated to the matter wave senses inertial or gravitational effects, the interferometer represents an accurate inertial probe.  
In particular, atom interferometers (AIs) have benefited from the outstanding developments of laser-cooling techniques and reached accuracies comparable to those of inertial sensors based on optical interferometry.
Thanks to their long term stability, AIs offer a breakthrough advance in accelerometry, gyroscopy and gravimetry, for applications to inertial guidance \cite{KasevichPatent}, geoid determinations \cite{Novak}, geophysics \cite{Imanishi} and metrology \cite{Geneves}. 

In addition, AIs are excellent candidates for laboratory-based tests of General Relativity (GR)  that could compete with the current tests which consider astronomical or macroscopic bodies \cite{Dimopoulos2007}. For example, AIs may provide new answers to the question of whether the free fall acceleration of a particle is universal, i.e. independent of its internal composition and quantum properties. Although this principle -- known as the Universality of Free Fall (UFF) -- has been tested experimentally \cite{Williams,Schlamminger} to a few parts in $10^{13}$, various extensions to the current theoretical physics framework predict its violation \cite{Will}. 
It is thus important to test experimentally these theoretical models with different types of particles.
AIs also open perspectives for further tests of GR such as the detection of gravitational waves \cite{Dimopoulos2008}. 
All these fundamental tests may benefit from the long interrogation times accessible on microgravity platforms \cite{Dimopoulos2007,Stern2009,Zoest} or in space \cite{Ertmer}.

Because of its high sensitivity, running an AI has required until now low vibration and high thermal stability environments which can only be found in dedicated ground or underground platforms. 
We report here the first operation of a matter wave inertial sensor in an aircraft, both at 1-g and in microgravity (0-g).
Our matter wave interferometer uses $^{87}\text{Rb}$ atoms and operates aboard the Novespace A300-0g aircraft taking off from Bordeaux airport, France \cite{Novespace}. 
This plane carries out parabolic flights during which 22 second  ballistic trajectories (0-g) are followed by two minutes of standard  gravity flight (1-g).  The AI measures the local acceleration of the aircraft with respect to the inertial frame attached to the interrogated atoms which are in free fall. 
In the first part of this communication, we describe the inertial measurements performed by our instrument and show how the matter wave sensor achieves a resolution level more than 300 times below the plane acceleration level. 
We present the general method that is used to operate the AI over a wide acceleration range  and to reach such a resolution.

In the second part, we demonstrate the first operation of a matter wave inertial sensor in 0-g.
Microgravity offers unique experimental conditions to carry out tests of fundamental physics.
However, these experiments are conducted on platforms such as planes, sounding rockets or satellites which are not perfectly free-falling, so that the residual craft vibrations might strongly limit the sensitivity of the tests.
Overcoming this problem generally requires the simultaneous operation of two sensors in order to benefit from a common mode vibration noise rejection. 
For example, conducting a matter wave UFF test implies the simultaneous interrogation of two different atomic species by two AIs  measuring their acceleration difference \cite{Fray04}. 
In the present work, we investigate the 0-g operation of our one-species AI in a differential configuration to illustrate a vibration noise rejection.
Our achievements (0-g operation and noise rejection) constitute major steps towards a 0-g plane based test of the UFF with matter waves at the 10$^{-11}$ level, and towards a space based   test \cite{Varoquaux} below $10^{-15}$. Such an experiment in space has been selected for the next medium-class mission in ESA's Cosmic Vision 2020-22 in the frame of the STE-QUEST project \cite{CosmicVision,SteQuest}.

This paper presents the first airborne and microgravity operation of a matter wave inertial sensor. We introduce a new and original method that allows to use the full resolution of an atom  interferometer in the presence of high levels of vibration. We also show how high precision tests of the weak equivalence principle may be conducted with differential atom interferometry.

\section*{Results}

\subsection*{Description of the airborne atom interferometer.}
Our experiment relies on the coherent manipulation of atomic quantum states using light pulses \cite{Borde,Kasevich1991}. We use telecom based laser sources that provide high frequency stability and power in a compact and integrated setup \cite{Lienhart}. 
Starting from a $^{87}\text{Rb}$ vapor, we load in $400 \ \text{ms}$ a cloud of about $3\times 10^{7}$ atoms laser cooled down to 10 $\mu$K, and select the atoms in a magnetic field insensitive ($m_{\text{F}}=0$) Zeeman sublevel.
We then apply a velocity selective Raman light pulse \cite{Kasevich2} carrying two counterpropagating laser fields so as to keep $10^{6}$ atoms which enter the AI with a longitudinal velocity distribution corresponding to a temperature of 300 nK. 
The Raman laser beams are aligned along the plane wings direction (Y axis, figure 1) and are retroreflected by a mirror attached to the aircraft structure and following its motion. 
The AI consists of a sequence of three successive Raman light pulses to split, redirect and recombine the atomic wavepackets (see figure 1D). 
The acceleration measurement process can be pictured as marking successive positions of the free-falling atoms with the pair of Raman lasers, and the resulting atomic phase shift $\Phi$ is the difference between the  phase of the two Raman lasers at the atom's successive classical positions, with respect to the retroreflecting mirror \cite{Storey}.
As the Raman beam  phase simply relates to the distance between the atoms and the reference retroreflecting mirror, the AI provides a measurement of the relative mean  acceleration $a_{\text{m}}$ of the mirror during the interferometer duration, along the Raman beam axis.
The information at the output  of the AI is a two-wave interference sinusoidal signal $P=P_0 -A\cos\Phi$, where $P$ is the transition probability between the two $^{87}\text{Rb}$ ground states, and $P_0$ (resp. $A$)  is the offset (resp. the amplitude) of the interference fringes.
This signal is modulated by the atomic phase shift $\Phi = a_{\text{m}}\times kT^2$, where $k = 2 \times 2\pi/780$ nm is the Raman lasers effective wavevector and $T$ is the time between the light pulses (see the first "Methods" subsection for the calculation of the phase shift).

\begin{figure}[!h]
\includegraphics[width=\linewidth]{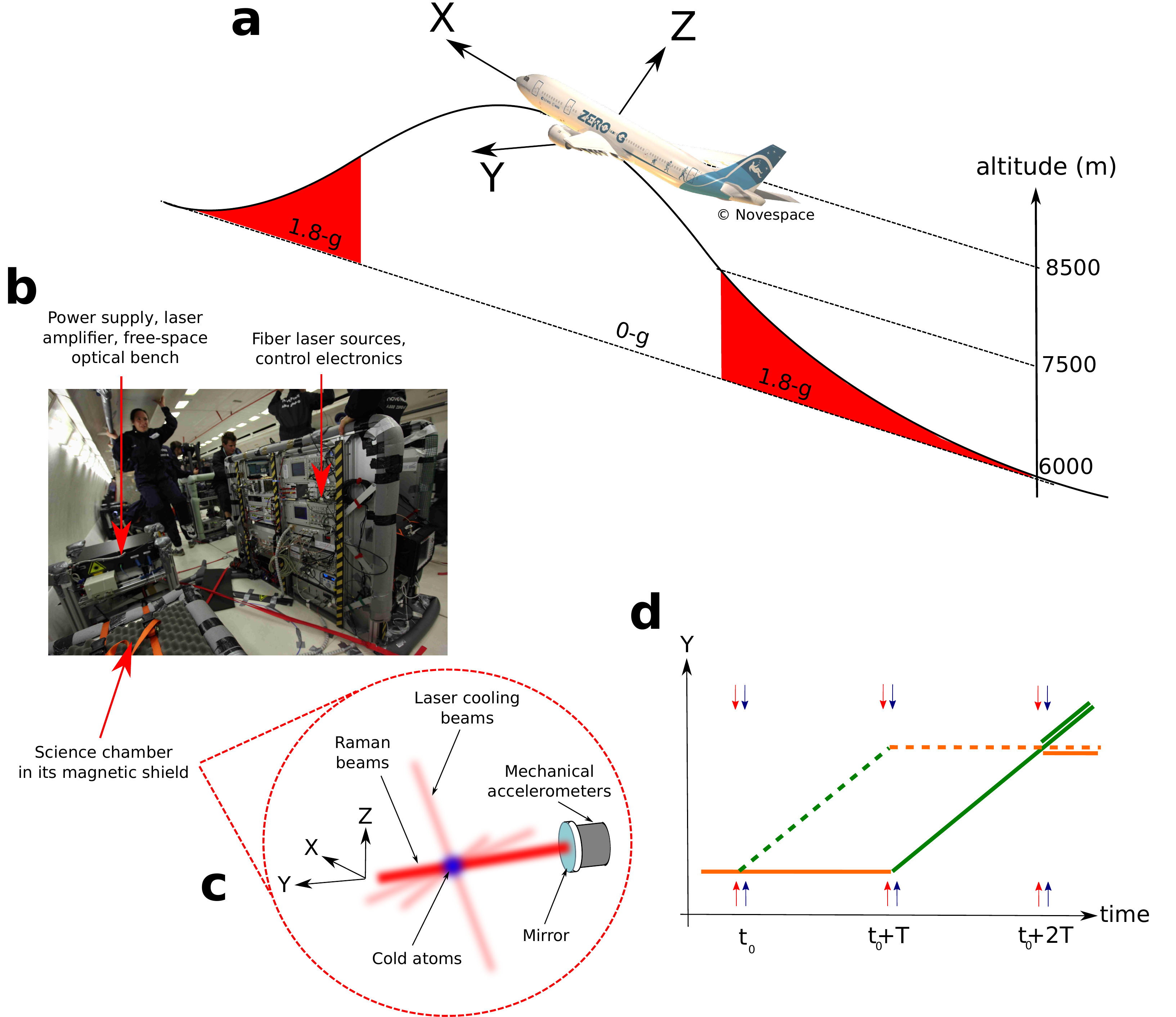}
\legend{\textbf{Fig. 1: Description of the experiment in the plane.} \textbf{(a)} The parabolic manoeuvre consists of a 20 second pull-up hypergravity (1.8-g) phase, the 22 second  ballistic trajectory (0-g) and a 20 second pull-out 1.8-g phase. 
This manoeuvre is alternated with standard gravity (1-g) phases of about two minutes and carried out 31 times by the pilots during the flight.
\textbf{(b)} Picture of the experiment in the plane during a 0-g phase.
 \textbf{(c)} Zoom in the science chamber where the atoms are laser cooled and then interrogated by the Raman laser beams (red) which are collinear to the Y axis and retroreflected by a mirror (blue). 
\textbf{(d)} Space-time diagram of the AI consisting of three successive light pulses which split, reflect and recombine the two matter waves represented by the dashed and the solid lines. The blue and red arrows represent the two Raman laser beams.}
\end{figure}

In the aircraft, the acceleration along Y (figure 2a) fluctuates over time by $\delta a_{\text{m}}\sim 0.5 \ \text{m.s}^{-2}$ (1 standard deviation), and is at least three orders of magnitude greater than the typical signal variations recorded by laboratory-based matter wave inertial sensors. For this reason, the signal recorded by the AI first appears as random, as shown in figure 2b. 
To quantify the information contained in the atomic measurements, we use mechanical accelerometers (MAs) fixed on the retroreflecting mirror and search for the correlation between the MAs and the AI \cite{Merlet2009}. 
We use the signal $a_{\text{MA}}(t)$ continuously recorded by the MAs to estimate the mean acceleration $a^{\text{E}}(t_i)$ which is expected to be measured by the AI at time $t_i=iT_c$, with $T_c=500 \ \text{ms}$  being the experimental cycle time (see the first "Methods" subsection). 
Plotting the atomic measurements $P(t_i)$ versus $a^{\text{E}}(t_i)$ reveals clear sinusoidal correlations between the mechanical sensors and the AI, both at 1-g (figure 2c) and in 0-g (figure 2d). 
This demonstrates that the AI truly holds information on the mirror acceleration $a_{\text{m}}$. 
We note that this result stands for the first demonstration of the operation of an atom accelerometer in an aircraft and in microgravity.

\begin{figure}[!h]
\includegraphics[width=\linewidth]{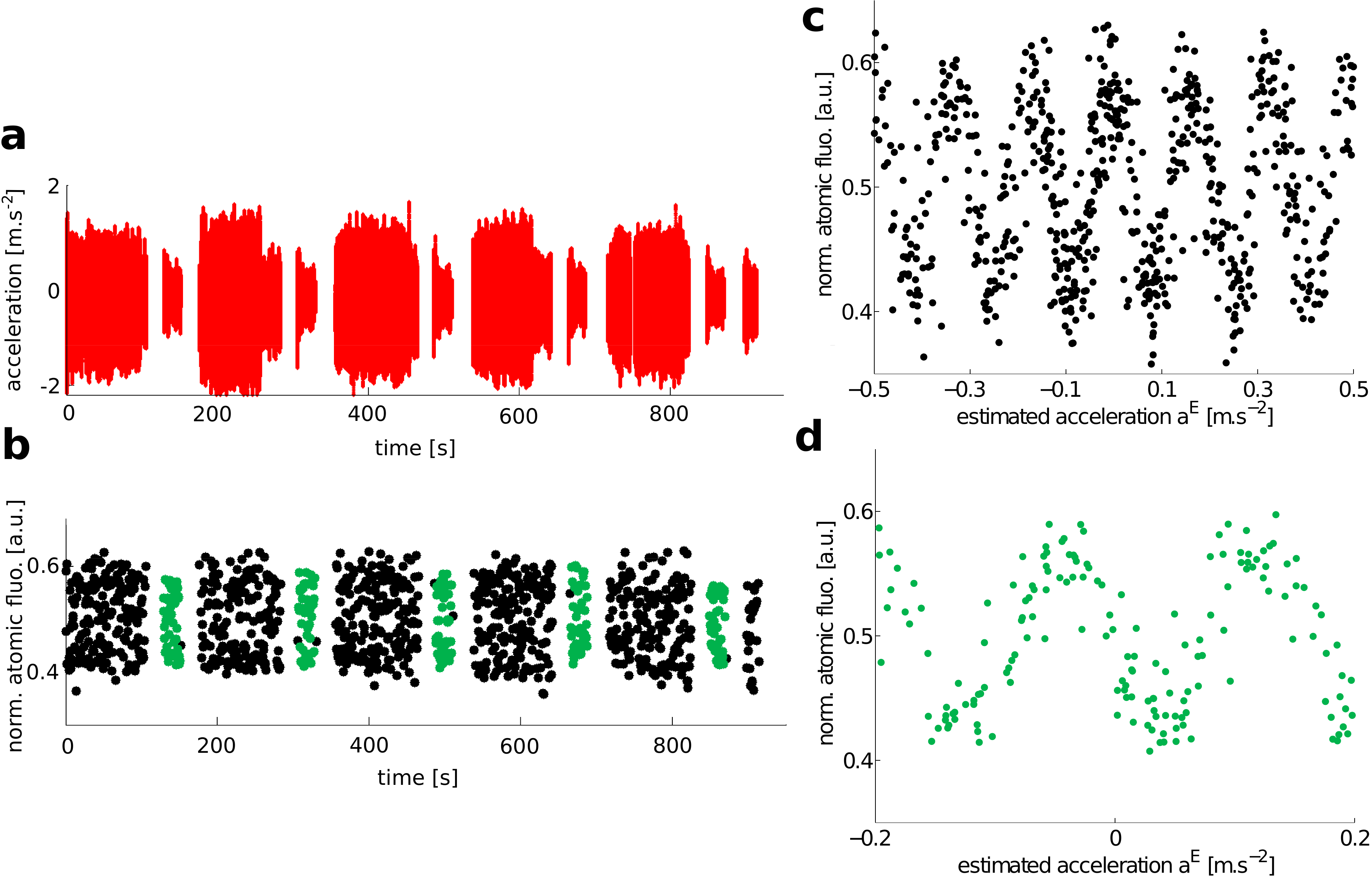}
\legend{\textbf{Fig. 2: AI revealing information on the plane acceleration.} \textbf{(a)} Acceleration signal recorded by the mechanical accelerometers (red); the standard deviation $\delta a_{\text{m}}$ of the acceleration signal is about 0.5 m.s$^{-2}$ at 1-g and 0.2 m.s$^{-2}$ in 0-g. \textbf{(b)} AI discrete measurements corresponding to the atomic fluorescence of the $^{87}\text{Rb}$ atoms in the $F=2$ state, normalized to the fluorescence of all the atoms; the total interrogation time is here $2T=3$ ms. The black and green points correspond to the 1-g and  0-g phases respectively; we have removed the 1.8-g phases where the AI is not designed to operate. \textbf{(c), (d)} Atomic measurements plotted versus the signal stemming from the mechanical accelerometers at 1-g (c) and in 0-g (d); the sinusoidal correlations show that the AI contains information on the acceleration of the plane.}
\end{figure}

\subsection*{Retrieving the plane acceleration with the AI resolution.}
We now consider the application of our matter wave sensor to precise measurements of the plane acceleration, by operating the AI beyond its linear range. 
For that purpose, we determine the AI acceleration response, defined by $P^{\text{AI}}(a_{\text{m}})=P_0 -A\cos(kT^2\times a_{\text{m}})$, independently from the mechanical devices.
We have developped a method (see the "Methods" section)  to estimate this response (i.e. the parameters $P_0$ and $A$), and the signal to noise ratio (SNR) of the interferometer, which determines the acceleration noise of the sensor, $\sigma_a = 1/(\text{SNR}\times kT^2)$.
The knowledge of $P_0$ and $A$ enables us to extract the acceleration $a_{\text{m}}(t_i)$ from  the atomic data $P(t_i)$ by inverting the model $P^{\text{AI}}(a_{\text{m}})$.
In this way, the acceleration is known within the region where the interferometer model can be inverted unambiguously and corresponding to an acceleration interval of range $a_{\text{R}}=\pi/kT^2$.
To obtain the total acceleration, we need the information  on the reciprocity region $[n(t_i) a_{\text{R}}, (n(t_i)+1)a_{\text{R}}]$ where the AI operates at measurement time $t_i$, with $n(t_i)$ being the interference fringe number where the measurement point is located (see the "Methods" section).
To determine $n(t_i)$, we use the MAs which have a reciprocal response over a wide acceleration range.
Thus, our instrument consists in a hybrid sensor which is able to measure large accelerations thanks to the mechanical devices, and able to reach a high resolution thanks to the atom accelerometer.
Figures 3a to 3c illustrate the measurement process that we use to measure the plane acceleration during successive 1-g and 0-g phases of the flight (figure 3d), with high resolution.

\begin{figure}[!h]
\includegraphics[width=\linewidth]{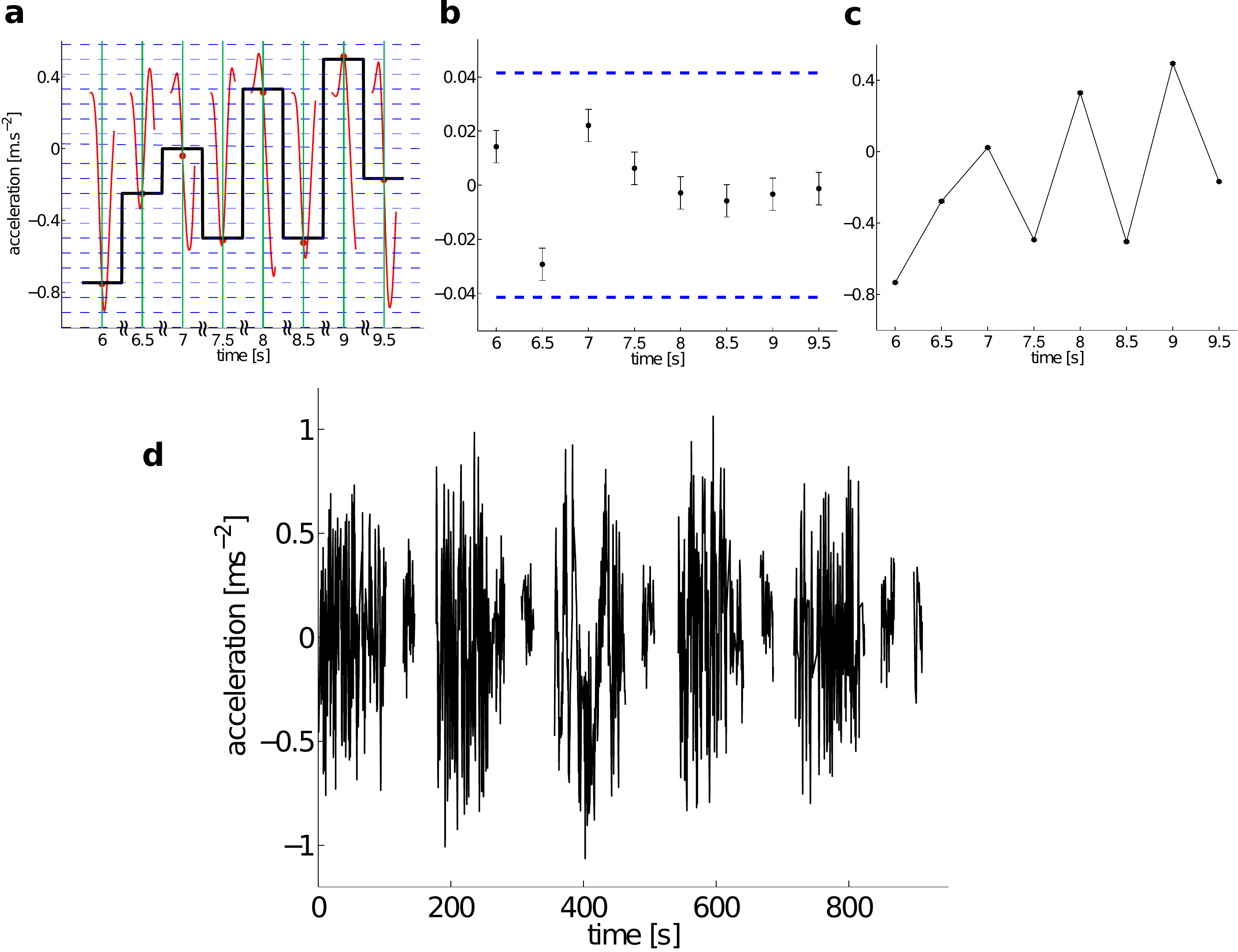}
\legend{\textbf{Fig. 3: High resolution measurement of the plane acceleration.} \textbf{(a)}  In red, signal recorded by the MAs in the time window $[-2T,2T]$ around the measurement times $t_i=iT_c$ (vertical green lines), with $T_c=500$ ms and $2T=3$ ms. The MAs signal has been filtered by the response function of the AI described in the "Methods" section, and the red points represent the value of the signal at $t_i$. This value determines the reciprocity region  where the AI operates, delimited by two horizontal dashed lines. 
In this way, the MAs provide the coarse acceleration measurement  (black step-like signal). \textbf{(b)} The AI is then used for the high resolution measurement within its reciprocity region, bounded by the two  blue dashed lines at $\pm a_{\text{R}}/2$, with $a_{\text{R}}=\pi/kT^2\approx 0.087 \ \text{m.s}^{-2}$. The error bars represent the  noise of the atom accelerometer, which equals $0.0065 \ \text{m.s}^{-2}$ per shot in this example ($\text{SNR}=4.3$). \textbf{(c)} The total acceleration ($a_{\text{m}}$) is the sum of the black step-like signal in (a) and of the AI measurements in (b). \textbf{(d)} Full signal measured by the hybrid MAs-AI sensor  aboard the A300-0g aircraft during successive 1-g and 0-g phases of the flight. For this data set, where $2T_{\text{MA}}=3 \ \text{ms}$, $\text{SNR}=4.3$ and $T_c=500 \ \text{ms}$, the  resolution of the sensor in one second is more than 100 times below the plane acceleration fluctuations $\delta a_{\text{m}}$.}
\end{figure}

\subsection*{Main error sources.}
Due to their limited performances, mainly their nonlinear response and intrinsic noise (see the "Methods" section), the MAs provide a signal which is not perfectly proportional to the acceleration $a_{\text{m}}$. 
This leads to errors on the estimated acceleration $a^{\text{E}}$  which blur the MAs-AI  correlation function and might prevent from finding the fringe index $n(t_i)$ where the AI operates.
These errors increase with the AI sensitivity and with the acceleration signal $a_{\text{m}}$.
For a given acceleration level, the good measurement strategy consists in increasing $T$ up to $T_{\text{MA}}$ where the MAs can still resolve the correlation fringes. This will set the scale factor $k T_{\text{MA}}^2$ of the AI. The sensitivity of the accelerometer is then determined by the SNR of the matter wave sensor, which is estimated independently from the MAs with our method. 
For instance at 1-g, where $\delta a_\text{m}\sim 0.5 \ \text{m.s}^{-2}$, the MAs enables us to increase the AI interrogation time up to $2T_{\text{MA}}=6 \ \text{ms}$  and to resolve the correlation fringes.
Future improvements will rely on the use of well characterized MAs, in particular on a calibration of their scale factor at the $10^{-3}$ level over the sensitivity bandwidth of the AI, to achieve interrogation times $2T_{\text{MA}}>20 \ \text{ms}$. 
High frequency ($>10 \ \text{Hz}$) vibration damping could also be used to constrain the frequency range where the MAs are needed, and therefore push forward a particular sensor technology.

Second, parasitic inertial effects due to the rotation of the plane might be experienced by the matter wave sensor and not by the MAs.
At 1-g, the atoms fall down before interacting with the Raman beams (aligned along the Y axis), so that the interferometer has a physical area and is thus sensitive to the Sagnac effect.
The resulting Coriolis acceleration measured by the AI and not by the MAs might impair the correlation and limit the performance of the hybrid sensor. 
For shot to shot fluctuations of the plane rotation of the order of $10^{-3} \ \text{rad.s}^{-1}$, we estimate a limit to the sensitivity at the $10^{-4} \ \text{m.s}^{-2}$ level at 1-g.
This error source may be significantly reduced in the future with the use of additional sensors to measure the rotation of the plane and  to take it into account in the calculation of the estimated acceleration.

Finally, the atomic SNR limits the sensitivity of the inertial sensor. 
During the flight, we measure at 1-g a SNR of 3.1 for $2T=6 \ \text{ms}$, which is in agreement with the value measured in our laboratory for the same interrogation time. 
With our experimental setup, the signal ($A\sim 0.1$) is essentially limited by the imperfections of the atomic beam splitters and mirror due to the temperature of the cloud and to the gaussian intensity profile of the Raman beams, while the noise is mainly due to detection noise.
In these conditions, with $T_c=500 \ \text{ms}$, the acceleration noise of the AI equals $1.6\times10^{-3} \ \text{m.s}^{-2}/\sqrt{\text{Hz}}$. 
At this sensitivity level, the hybrid sensor is able to measure inertial effects more than 300 times weaker than the typical  acceleration fluctuations of the aircraft.
We emphasize that reaching such a high resolution is possible thanks to the appropriate combination of MAs (see the "Methods" section), the success of operating the AI in the plane, and the use of our method for the acceleration measurement.
In the present configuration of the experiment, the SNR degrades at 1-g when $2T$ increases above $20 \ \text{ms}$ because the atoms fall down and escape the Raman and detection beams  \cite{Stern2009}. 
This limitation could be overcome thanks to additional Raman beam collimators and by changing the orientation of the detection lasers.
In 0-g, the experiment falls with the atoms and the SNR is not constrained by gravity any more.
The SNR may also be improved significantly in the future thanks to a better detection system (e.g. more stable detection lasers) and the use of ultra-cold atoms.

\subsection*{Differential measurement in 0-g.}
We focus now on the microgravity operation of the matter wave inertial sensor and on its possible application to fundamental physics tests such as that of the UFF.
Such a test can be carried out with two AIs measuring the acceleration of two different atoms  with respect to the same mirror that retroreflects the Raman lasers. 
In airborne or spaceborne experiments, the mirror constitutes an ill-defined inertial reference because of the craft's vibrations.
This might degrade the sensitivity of the test, unless the vibrations  impact the two interferometers in the same way. 

Vibration noise rejection occurs when conducting differential measurements such as in the operation of gradiometers \cite{McGuirk} or gyroscopes \cite{Gauguet}, and is expected in UFF tests based on atom interferometry \cite{Varoquaux}.
The rejection efficiency  depends on the two species used for the UFF test and is maximum for simultaneous interrogation of the two atoms (see supplementary information).
In the case of a finite rejection, the  impact of the acceleration noise can further be reduced by measuring the vibrations of the mirror with MAs to substract them from the differential phase measurement.
In this way, the MAs are used to release the requirements on the vibration damping of the craft as they measure the accelerations not rejected in the differential operation of the two AIs. 
The efficiency of that technique is limited by the performances of the MAs as their imperfections (e.g. their non-linearities) translate into residual vibration noise impairing the differential acceleration measurement.
In the following, we investigate  vibration noise rejection by operating our $^{87}$Rb sensor in a differential mode.
We use a sequence of 4 light pulses to build a two-loop AI \cite{McGuirk} which is  equivalent to two successive one-loop interferometers head to tail (figure 4b). 
The 4-pulse AI provides a signal resulting from the coherent substraction of two spatially and temporally separated inertial measurements and is therefore expected to be less sensitive to the low frequency inertial effects.

\begin{figure}[!h]
\centering
\includegraphics[width=\linewidth]{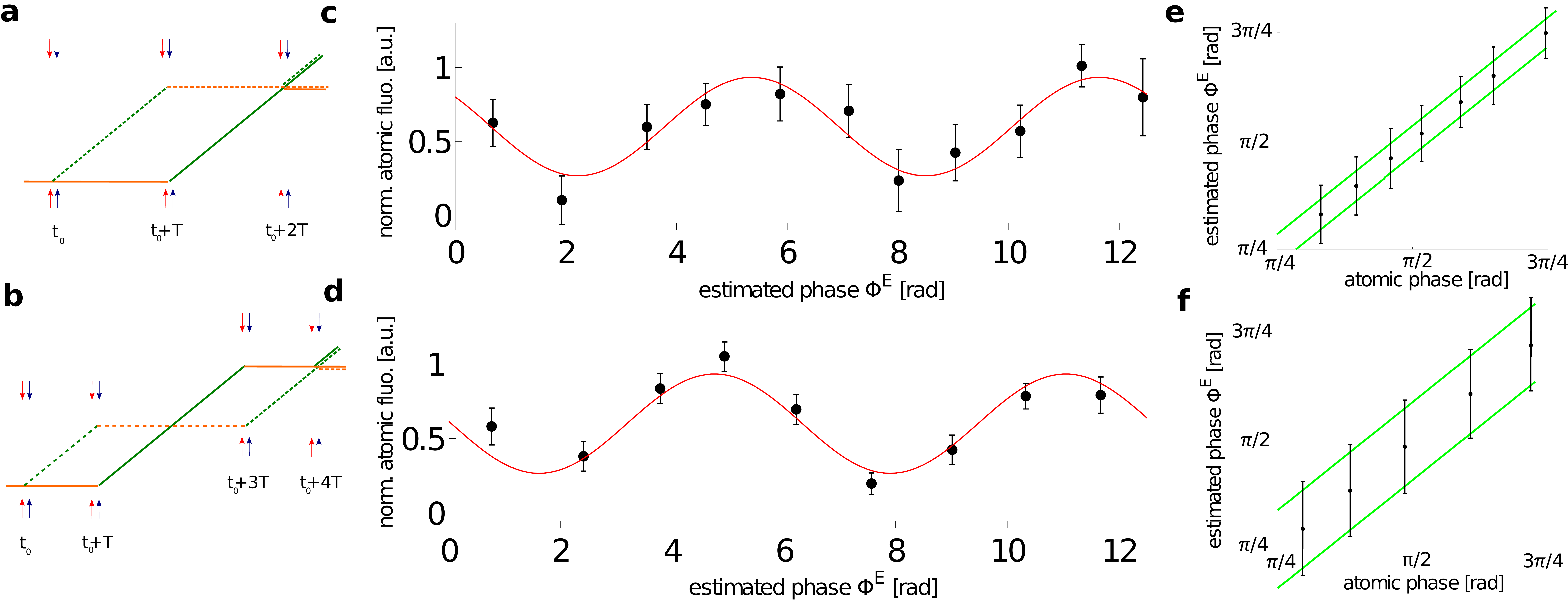}
\legend{\textbf{Fig. 4: Comparison of two AI geometries in 0-g.} \textbf{(a), (b)} 3-pulse and 4-pulse interferometers considered in this work. \textbf{(c), (d)} Corresponding MAs-AI correlation functions recorded  during the 0-g phase of consecutive parabolas, with total interrogation times of $2T=$ 20 ms and  $4T=$ 20 ms,  respectively. To obtain these plots, we have sorted the estimated phase data (200 and 180 points respectively) and averaged the correlation points by packets of 20. The error bars equal the standard deviation of each packet divided by $\sqrt{20}$, and are transferred to the vertical axis in this averaging procedure. The red line is a sinusoidal fit to the points. To make the comparison of the two correlations easier, we have scaled the vertical axis so as to obtain the same amplitude for the two sinusoids. \textbf{(e), (f)} Comparison of the total correlation noise $\sigma_{\text{corr}}$ (error bars) and of the atomic phase noise $\sigma_{\text{AI}}= 1/(\text{SNR}\times\sqrt{20})$  for the two interferometer geometries (the vertical spacing between the green lines is $2\sigma_{\text{AI}}$). The atomic phase data $\{\Phi(t_i)\}$ have been obtained from the atomic measurements $\{P(t_i)\}$ by inverting the AI model $P^{\text{AI}}(\Phi)$, in the linear region of the interferometer ranging from $\Phi\approx \pi/4$ to $\Phi\approx 3\pi/4$. To facilitate the comparison of $\sigma_{\text{corr}}$ and $\sigma_{\text{AI}}$, we have  set $\Phi^{\text{E}}=\Phi$  (the points are thus aligned on the first bisector). Figure (e) (resp. (f)) shows that  the sensitivity of the sensor is limited by the vibration noise $\sigma_{\text{vib}}$  not measured by the MAs (resp. by the atomic phase noise).}
\end{figure}

To illustrate the noise rejection, we  operate the 3-pulse and 4-pulse interferometers  with a same total interrogation time of 20 ms,  and  compare the MAs-AI correlation for each geometry (figure 4). 
Figures 4c and 4d show that the quality of the correlation (ratio of the sinusoid contrast  to the mean error bar) is clearly improved in the differential geometry compared to the 3-pulse interferometer.
To understand this difference, we estimate the total noise of the correlation, defined as  $\sigma_{\text{corr}}=\sqrt{\sigma_{\text{AI}}^2+\sigma_{\text{vib}}^2}$. 
It results, in the linear range of the interferometer, from the quadratic sum of two independent contributions: the atomic phase noise $\sigma_{\text{AI}}\propto 1/\text{SNR}$, and the  vibration noise $\sigma_{\text{vib}}$  not measured by the MAs.
As we can estimate $\sigma_{\text{AI}}$ independently from the correlation (see the second "Methods" subsection), the comparison of $\sigma_{\text{corr}}$ and $\sigma_{\text{AI}}$ indicates whether the sensor sensitivity is limited by the atomic noise ($\sigma_{\text{corr}}\approx \sigma_{\text{AI}}$) or by the residual vibration noise ($\sigma_{\text{corr}} > \sigma_{\text{AI}}$).
In figures 4e and 4f, we have represented $\sigma_{\text{corr}}$ (errors bars) and $\sigma_{\text{AI}}$  (vertical spacing between the green lines) for the one-loop and the two-loop  interferometers, respectively. 
Figure 4e shows that the sensitivity of the 3-pulse sensor is limited by the vibration  noise not measured by the MAs, as $\sigma_{\text{corr}} > \sigma_{\text{AI}}$.
On the contrary, figure 4f reveals  that the atomic phase noise  is the main limit to the sensitivity of the 4-pulse sensor, as $\sigma_{\text{corr}}\approx \sigma_{\text{AI}}$.
The SNR in the 4-pulse interferometer is less than in its 3-pulse counterpart as the additional  light pulse reduces the interference fringe contrast due to the Raman beam intensity inhomogeneities and the transverse temperature of the cloud.
In spite of greater atomic noise, the quality of the correlation 4f is better than that of  correlation 4e, which shows that the 4-pulse sensor operates a vibration noise rejection.

In the 4-pulse geometry, the two elementary interferometers operate one after another and share accelerations of frequency below $1/2T$.
In a UFF test, two AIs of equal scale factor ($kT^2$) will interrogate two different atoms almost simultaneously, so that the noise rejection is expected to be much more efficient (see \cite{Varoquaux} and the supplementary information). 
In that case, the precision of the inertial sensor might be limited by the atomic phase noise, i.e. by the SNR of the interferometer.
For the 3-pulse AI data in figure 4c and 4e ($2T=20 \ \text{ms}$), we estimate a SNR of 2.1, which corresponds to an acceleration sensitivity of the matter wave sensor of $2\times 10^{-4} \ \text{m.s}^{-2}/\sqrt{\text{Hz}}$, in the context of the high vibration noise rejection expected for the UFF test. 
That sensitivity level may be greatly improved in the future by using a highly collimated atomic source in microgravity \cite{Zoest}.

\section*{Discussion}
We finally discuss possible improvements of our setup, both for inertial guidance and fundamental physics applications.
In the former case, increasing the resolution of the accelerometer will be achieved by using  well characterized MAs to increase $T_{\text{MA}}$, and by improving the SNR of the AI.
Reaching an interrogation time $2T_{\text{MA}}=40 \ \text{ms}$ in the plane (where the acceleration  fluctuations are $\delta a_{\text{m}}\sim0.5 \ \text{m.s}^{-2}$ rms at 1-g) would require MAs whose scale factor frequency response is determined with a relative accuracy of $2\times 10^{-4}$. We believe that such precision can be achieved with state-of-the-art MA technology, e.g. capacitive MEMS sensors, and we now work at implementing these sensors in our setup.
Together with a shot noise limited  AI  with $\text{SNR}\sim 200$ (as in \cite{Gauguet}), the resolution of the hybrid sensor would be $8\times10^{-7} \ \text{m.s}^{-2}$ per shot, which would represent a major advance in inertial navigation as well as in airborne gravimetry.
We note that the present analysis does not report the in-flight bias of the   matter wave sensor since no other airborne accelerometer of similar accuracy   was available onboard to proceed to the comparison. However, it has been   demonstrated in laboratories that atom accelerometers can reach biases of the   order of $10^{-8} \ \text{m.s}^{-2}$ under appropriate conditions \cite{Merlet2010}.

Tests of fundamental physics would also require performant MAs to remove the residual  aircraft's or satellite's acceleration noise not rejected in the differential measurement.
For a UFF test in the 0-g plane and a vibration rejection efficiency of 300 (explained in \cite{Varoquaux} and in the supplementary information),  a differential acceleration sensitivity of $3\times 10^{-10} \ \text{m.s}^{-2}$ per shot  ($\text{SNR}= 200$, $2T=2 \ \text{s}$)  could be achieved with MAs of $2\times 10^{-4}$ relative accuracy  if the vibrations during the 0-g phase are  damped to the $5\times 10^{-4} \ \text{m.s}^{-2}$ level.
In space, high performance MAs such as the sensors developped for the GOCE mission could be used to determine the residual accelerations of the satellite ($\sim 10^{-6} \ \text{m.s}^{-2}$) with a resolution \cite{Marque2008} of the order of $ 10^{-12} \ \text{m.s}^{-2}$. 
The vibration rejection efficiency of 300  would thus limit the impact of the acceleration noise on the interferometric measurement to $3\times 10^{-15} \ \text{m.s}^{-2}$ per shot, which would  stand for a minor contribution in the error budget.
Therefore, high precision test of the equivalence principle could be conducted in space without the strong drag-free constraints on the satellite that represent a major challenge in current space mission proposals.

To conclude, we have demonstrated the first airborne operation of a cold atom inertial sensor, both at 1-g and in microgravity. 
We have shown how the matter wave  sensor can measure the craft acceleration with high resolutions. Our approach proposes to use mechanical devices which probe the coarse inertial effects and allow us to enter the fine measurement regime provided by the atom accelerometer. 
In the future, instruments based on the combination of better characterized mechanical sensors  and a shot noise limited  AI  could reach sensitivities of the order of few $10^{-7} \ \text{m.s}^{-2}$ in one second aboard aircrafts.
Thus, our investigations indicate that sensors relying on cold atom technology may be able to detect inertial effects with resolutions unreached so far by instruments aboard moving crafts characterized by high acceleration levels. 
Cold atom sensors offer new perspectives in inertial navigation thanks to their long term stability as they could be used to correct the bias ($\sim \ \text{few} \ 10^{-5} \ \text{m.s}^{-2}$) of the traditional sensors monitoring the crafts' motion, below the $10^{-7} \ \text{m.s}^{-2}$ level.
In geophysics, airborne gravity surveys may also benefit from the accuracy of AIs \cite{Wu}. 

Moreover, we have operated the first matter wave sensor in microgravity. 
We have shown how differential interferometer geometries enable to reject vibration noise of the experimental platform where new types of fundamental physics tests will be carried out.
Our result in 0-g suggests that the high sensitivity level of matter wave interferometers may be reached on such platforms, and support the promising future of AIs to test fundamental physics laws  aboard  aircrafts, sounding rockets or satellites  where long interrogation times can be achieved \cite{CosmicVision}. 
While many quantum gravity theories predict violations of the UFF, AIs may investigate its validity at the atomic scale, with accuracies comparable to those of ongoing or future  experiments monitoring macroscopic or astronomical bodies \cite{Touboul}.


\section*{Methods}

\subsection*{AI response function and  MAs-AI correlation.}
For a time varying acceleration $a(t)$ of the retroreflecting mirror, the phase of the interferometer at time $t_i=iT_c$ is given by
\begin{equation}
\Phi(t_i)=k\int f(t,\tilde{t_i})a(t)dt,
\end{equation}
where $f$ is the acceleration response of the 3-pulse AI. It is a triangle-like function \cite{Varoquaux}  that reads:
\begin{equation}
f(t,\tilde{t_i}) = \left\{ \begin{array}{ll}
 t-\tilde{t_i} & \text{if} \ t \in [\tilde{t_i},\tilde{t_i}+T] \\
 2T-(t-\tilde{t_i}) & \text{if} \ t \in [\tilde{t_i}+T,\tilde{t_i}+2T],
\end{array} \right.
\end{equation} 
with  $\tilde{t_i}$ being the time of the first Raman pulse at the $i$-th measurement. 
The mean  acceleration that we infer is defined as $a_{\text{m}}=\Phi/k T^2$.

Thanks to the MAs, we estimate the phase $\Phi^{\text{E}}(t_i)$ which is expected to be measured by the interferometer by averaging in the time domain the signal $a_{\text{MA}}(t)$ by the AI response function:
\begin{equation}
\Phi^{\text{E}}(t_i)=k\int f(t,\tilde{t_i})a_{\text{MA}}(t)dt.
\end{equation}
The MAs-AI correlation function can be written as
\begin{equation}
\label{eq:corr}
P=P_0 -A\cos\Phi^{\text{E}},
\end{equation}
and expresses the probability to measure the atomic signal $P(t_i)$ at time $t_i$, given the acceleration signal $a_{\text{MA}}(t)$ recorded by the MAs. The estimated acceleration used in figure 2 is defined by $a^{\text{E}}=\Phi^{\text{E}}/kT^2$.

For simplicity, we have neglected in equation (1) the Raman pulse duration $\tau= 20 \ \mu s$ with respect to the interrogation time $2T$.
The exact formula can be found in \cite{CheinetPhD} and has been used in the data analysis to estimate the phase $\Phi^{\text{E}}$.

\subsection*{Estimation of the AI response and signal to noise ratio.}
We calculate the Probability Density Function (PDF) of the AI measurements $P(t_i)$ and fit it with the PDF of a pure sine (a "twin-horned" distribution) convolved with a gaussian of standard deviation $\sigma_P$. 
The fit function reads:
\begin{equation}
\mathcal{F}(x)=\mathcal{N}\int_{-\infty}^{+\infty} dx^{\prime}\Big[1-\Big(\frac{P_0-x^{\prime}}{A}\Big)^2\Big]^{-1/2}  \times \frac{1}{\sigma_P\sqrt{2\pi}}\text{exp}\Big({-\frac{(x-x^{\prime})^2}{2\sigma_P^2}}\Big),
\end{equation}
where $\mathcal{N}$ is a normalization factor.
In this way, we estimate the amplitude $A$ and the offset value $P_0$ of the interference fringes, i.e. we estimate the AI response $P^{\text{AI}}(\Phi)=P_0 -A\cos\Phi$. The fitted parameters $A$ and $\sigma_P$ provide an estimate of the in-flight Signal to Noise Ratio of the interferometer, given by $\text{SNR} = A/\sigma_P$. 
Figure 5 illustrates the method for the data used in figure 2 and 3, corresponding to the interrogation time $2T=3$ ms. 
The fitted parameters are $P_0=0.50$, $A=0.074$ and SNR$=4.3$. 

\begin{figure}[!h]
\centering
\includegraphics[width=\linewidth]{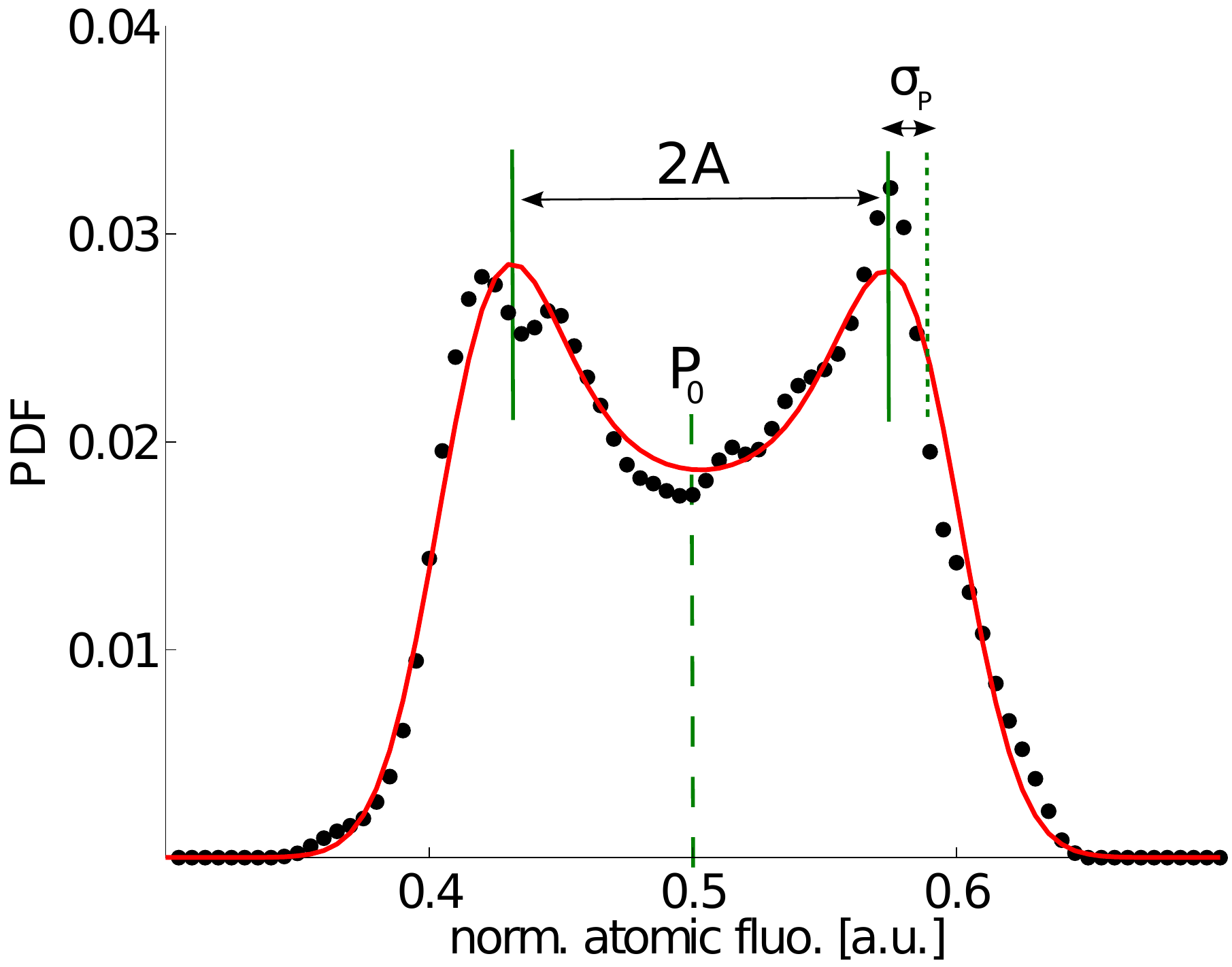}
\legend{\textbf{Fig. 5: Estimation of the AI response and  SNR.} Probability Density Function (PDF) of the AI measurements $P(t_i)$ (normalized atomic fluorescence) for the data of figure 2 and 3.  
$P_0$ and $A$ are respectively the offset and the amplitude of the interference fringes, and $\sigma_P$ is the standard deviation of the atomic noise.}
\end{figure}

This method is independent on the MAs signal and informs both on the response of the AI, and on its one-shot acceleration sensitivity $\sigma_a=1/(\text{SNR}\times kT^2)$.
The characteristics of the interferometer are thus known without any additional calibration procedure.
In particular, we can estimate the noise level of the matter wave sensor, given by $\sqrt{T_c}\sigma_a$ for a white atomic phase noise, and which equals $4.6\times 10^{-3} \ \text{m.s}^{-2}/\sqrt{\text{Hz}}$ for the data in figure 5, where $T_c=500 \ \text{ms}$.

Our analysis estimates  the SNR by taking into account only the atomic noise due to detection noise or fluctuations of the fringe offset and contrast.
It does not account for the laser phase noise that could impact the sensitivity of the acceleration measurement for long interrogation times \cite{Nyman06}. 
However, we demonstrated  the low phase noise of our laser system  during previous parabolic flight campaigns \cite{Stern2009}, which is at least one order of magnitude below the estimated atomic phase noise for the interrogation times we consider in this work ($T\le10 \ \text{ms}$).
Therefore, we have neglected in this communication the laser phase noise contribution when evaluating the sensitivity of the sensor.

\subsection*{Determination of the acceleration signal.}
Thanks to the MAs signal, we first determine the fringe number $n(t_i)=\text{floor}[a^{\text{E}}(t_i)/a_{\text{R}}]$ where the interferometer operates at time $t_i=iT_c$, with $a_{\text{R}}=\pi/kT^2$ being the reciprocity interval of the matter wave sensor. The values $n(t_i)$ are represented by the black step-like  curve in figure 3a.
Second, we use the atomic measurements $P(t_i)$ to deduce the acceleration $\tilde{a}(t_i)$ measured by the AI in its reciprocity region $[n(t_i)a_{\text{R}},(n(t_i)+1)a_{\text{R}}]$, and given by:
\begin{equation}
\tilde{a}(t_i)=\frac{1}{kT^2}\arccos\Big(\frac{P_0-P(t_i)}{A}\Big).
\end{equation}
The total acceleration $a_{\text{m}}(t_i)$  is finally computed as:

\begin{equation}
a_{\text{m}}(t_i) = \left\{ \begin{array}{ll}
 n(t_i) a_{\text{R}} + \tilde{a}(t_i) & \text{if} \ n(t_i) \ \text{is even} \\
 (n(t_i)+1)a_{\text{R}} - \tilde{a}(t_i) & \text{if} \ n(t_i) \ \text{is odd}.
\end{array} \right.
\end{equation}
The acceleration measurement process is illustrated in figure 6.

\begin{figure}[!h]
\includegraphics[width=1\linewidth]{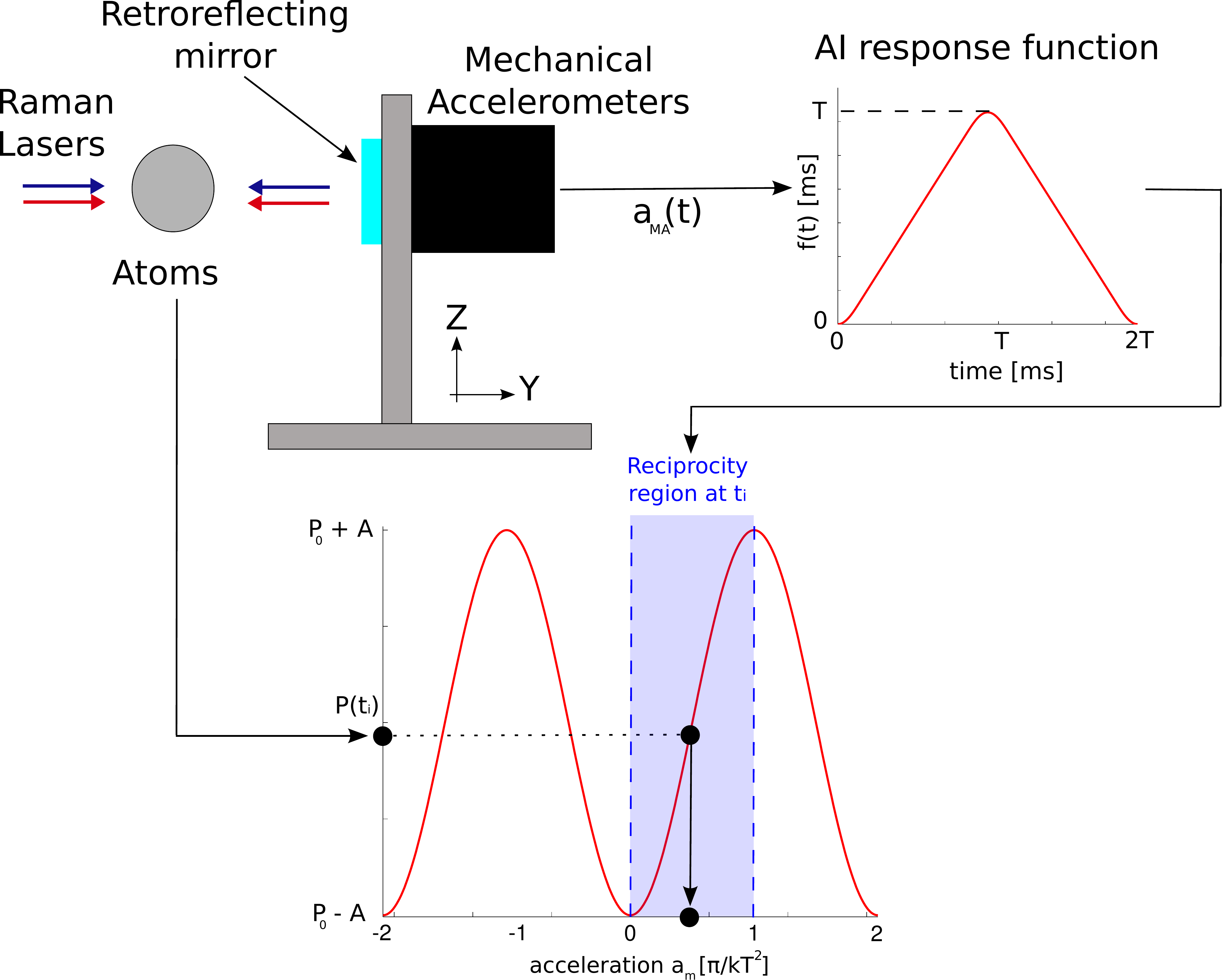}
\legend{\textbf{Fig. 6: Schematic of the two-step acceleration measurement process.} The signal of the MAs is filtered by the response function $f(t)$ of the AI (described in the first paragraph of the "Methods" section) and informs on the reciprocity region where the AI operates at the measurement time $t_i=i T_c$. In this example, the reciprocity region corresponds to the $0-\pi/kT^2$ interval, i.e. to $n(t_i)=0$. The value provided by the AI, $P(t_i)$, is then used to refine the acceleration measurement within the reciprocity region. The acceleration $\tilde{a}(t_i)$  is obtained from $P(t_i)$ by inverting the AI response $P^{\text{AI}}(a_{\text{m}})$ (red curve).}
\end{figure}

\subsection*{Limitations due to the Mechanical Accelerometers (MAs).}
The main limitation of the MAs comes from the nonlinearities in their frequency response (phase and gain), which means that the signal $a_{\text{MA}}(t)$ is not exactly proportional to the acceleration of the retroreflecting mirror. This results in errors in the estimation of the phase $\Phi^{\text{E}}$ that impair the MAs-AI correlation.
The nonlinearities typically reach amplitudes $\epsilon_{\text{nl}}\sim5\%$ within the AI bandwidth which equals  $1/2T\le$ 500 Hz. (The acceleration frequency response $H(\omega)$ of an AI has been measured in \cite{Cheinet08} and corresponds to the response of  a second order low pass filter of cutoff frequency $1/2T$). 
To reduce them, we combine two mechanical devices of relatively flat frequency response within two complementary frequency bands: a capacitive accelerometer (Sensorex SX46020) sensing the low frequency accelerations (0 to 1 Hz), and a piezoelectric sensor (IMI 626A03) measuring the rapid fluctuations (1 to 500 Hz). 
In this way, we achieve $\epsilon_{\text{nl}}\sim 2\%$. For $2T_{\text{MA}}=6 \ \text{ms}$, these nonlinearities correspond to errors  on $\Phi^{\text{E}}$ of about 0.5 rad rms. 
Further details on the errors  in the estimation of the phase due to the MAs nonlinearities are given in the  Supplementary Information.

Another limitation comes from the MAs internal noise, especially that of the capacitive one whose noise level integrated in [0-1Hz] equals $3\times 10^{-4} \ \text{m.s}^{-2}$. 
This noise is about one order of magnitude lower than this due to the non-linearities of the MAs, and is independent of the acceleration level in the plane.
Axis cross-talk of the MAs of the order of $2\%$ is taken into account for the estimation of the phase, so that the errors on $\Phi^{\text{E}}$ due to the MAs axis coupling are negligible.
Finally, the bias of the capacitive accelerometer is specified at the $0.05 \ \text{m.s}^{-2}$ level, so that the mean estimated acceleration might drift during the flight (4 hours).
This results in a displacement of the dark fringe position in the MAs-AI correlation plots.

\section*{Acknowledgments}
 This work is supported by CNES,  DGA, RTRA-Triangle de la Physique, ANR, ESA, ESF program EUROQUASAR and FP7 program iSENSE.
Laboratoire Charles Fabry and LNE-SYRTE are part of IFRAF.
We are grateful to Alain Aspect and Mark Kasevich for discussions and careful reading of the manuscript, and to Linda Mondin for her productive implication in the I.C.E. project. 
We thank  Andr\'e Guilbaud and the Novespace staff for useful technical advice.



\end{document}